\documentclass[showpacs,preprintnumbers,
amsmath,amssymb,aps,prd]{revtex4}
\usepackage{amsfonts}
\usepackage{amsfonts,graphics,epsfig,subfigure}

\begin{document}

\title{Combined effects of f(R) gravity and conformally invariant Maxwell field on the extended phase space thermodynamics of higher-dimensional black holes}
\author{Jie-Xiong Mo \footnote{mojiexiong@gmail.com}, Gu-Qiang Li \footnote{zsgqli@hotmail.com}, Xiao-Bao Xu \footnote{xbxu789@163.com}}

 \affiliation{Institute of Theoretical Physics, Lingnan Normal University, Zhanjiang, 524048, Guangdong, China}

\begin{abstract}
In this paper, we investigate the thermodynamics of higher-dimensional $f(R)$ black holes in the extended phase space. Both the analytic expressions and numerical results for the possible critical physical quantities are obtained. It is proved that meaningful critical specific volume only exists when $p$ is odd. This unique phenomenon may be attributed to the combined effect of $f(R)$ gravity and conformally invariant Maxwell field. It is also shown that the ratio $P_cv_c/T_c$ differs from that of higher dimensional charged AdS black holes in Einstein gravity. However, the ratio for four-dimensional $f(R)$ black holes is the same as that of four-dimensional RN-AdS black holes, implying that $f(R)$ gravity does not influence the ratio. So the ratio may be related to conformally invariant Maxwell field. To probe the phase transition, we derive the explicit expression of the Gibbs free energy with its graph plotted. Phase transition analogous to the van der Waals liquid-gas system take place between the small black hole and the large black hole. Classical swallow tail behavior, characteristic of first order phase transition, can also be observed in the Gibbs free energy graph. Critical exponents are also calculated. It is shown that these exponents are exactly the same as those of other AdS black holes, implying that neither $f(R)$ gravity nor conformally invariant Maxwell field influence the critical exponents. Since the investigated black hole solution depends on the form of the function $f(R)$, we discuss in detail how our results put constraint on the form of the function $f(R)$ and we also present a simple example.

\end{abstract}


  \maketitle

\section{Introduction}
    Black holes can be viewed as thermodynamic systems because they not only have temperature and entropy, but also exhibit rich critical phenomena. The famous Hawking-Page phase transition\cite{Hawking2} was found to take place between large AdS black hole and thermal AdS space. It was also discovered that charged AdS black holes undergo first-order phase transition analogous to the Van der Waals (VdW) liquid-gas system~\cite{Chamblin1,Chamblin2}. Recently, Kubiz\v{n}\'{a}k and Mann~\cite{Kubiznak} investigated the $P-V$ criticality of charged AdS black holes in the extended space where the cosmological constant is identified as thermodynamic pressure. Their enlightening research further enhanced the analogy between charged AdS black holes and liquid-gas systems. Moreover, reentrant phase transitions~\cite{Gunasekaran,Altamirano1,Altamirano3,Shaowen1,Frassino,Hennigar} reminiscent of multicomponent liquids and Small/intermediate/large black hole phase transitions~\cite{Frassino,Altamirano2,Shaowen1} reminiscent of solid/liquid/gas phase transition were discovered. These findings revealed the amazingly close relation between AdS black holes and ordinary thermodynamic systems. So far there has been so much effort concentrating on the extend phase space thermodynamics of black holes~\cite{Kastor}-\cite{Lee} that we can not list all of them here. For nice reviews, see Ref.~\cite{Altamirano3,Dolan4} and references therein.

     Among the research above, Chen et al.~\cite{Chen} investigated $P-V$ criticality of four-dimensional AdS black hole in $f(R)$ gravity. It was shown that $f(R)$ correction influences the Gibbs free energy and the ratio $\rho_c$. Recently, we further studied the coexistence curve and molecule number density of four-dimensional $f(R)$ AdS black holes~\cite{xiong5}. Specifically, we derived the analytic expressions of the universal coexistence curve that is independent of theory parameters and obtained the explicit expressions of the physical quantity describing the difference of the number densities of black hole molecules between the small and large black hole.

     However, extended phase space thermodynamics of higher-dimensional $f(R)$ AdS black holes has not been reported in literature yet. The attempt to obtain higher-dimensional black hole solutions in $f(R)$ gravity coupled to standard Maxwell field failed because in higher dimensions the standard Maxwell energy-momentum tensor is not traceless. Sheykhi creatively considered conformally invariant Maxwell action instead and successfully solved this problem. Both the black hole solutions and thermodynamics were thoroughly investigated~\cite{Sheykhi}. In this paper, we would like to generalize this research to the extended phase space where the cosmological constant is view as a variable and identified as thermodynamic pressure. The motivation is as follows. On the one hand, studying the black holes and their thermodynamics~\cite{Moon98}-\cite{Mazharimousavi} in $f(R)$ gravity has its own right. $f(R)$ gravity, as a kind of modified gravity theory, serves as an alternative approach other than dark energy to explain the cosmic acceleration. It is natural to expect that the thermodynamics of black holes in $f(R)$ gravity distinguishes itself from the thermodynamics of black holes in Einstein gravity. Moreover, the extended phase space thermodynamics of higher-dimensional black holes may reveal unique features associated with conformally invariant Maxwell field and higher-dimensional spacetime different from that of four-dimensional black holes. On the other hand, the mass should be interpreted as enthalpy rather than internal energy in the extended phase space. And the Smarr relation matches exactly with the first law of black hole thermodynamics. As stated in the first paragraph, the study of extended space thermodynamics has scored great success. Probing the thermodynamics of higher-dimensional $f(R)$ black holes in the extended phase space will further deepen the understanding of the relation between AdS black holes and ordinary thermodynamic systems.

    The organization of this paper is as follows. A short review on the thermodynamics of higher-dimensional $f(R)$ black holes will be presented in Sec. \ref{Sec2}. Then we will generalize the thermodynamics to the extended phase space in Sec. \ref{Sec3} and study the critical exponents in Sec. \ref{Sec4}. In Sec. \ref{Sec5}, we will discuss in detail how our results put constraint on the form of the function $f(R)$ and we will also present a simple example. A brief conclusion will be drawn in Sec. \ref {Sec6} .

\section{A brief review on the thermodynamics of higher-dimensional f(R) black holes}
\label{Sec2}
The approach to obtain higher-dimensional black hole solutions from $R+f(R)$ gravity coupled to standard Maxwell field failed because in higher dimensions the standard Maxwell energy-momentum tensor is not traceless. To solve this problem, one should consider the following conformally invariant Maxwell action instead
\begin{equation}
S_m=-\int d^nx\sqrt{-g}(F_{\mu\nu}F^{\mu\nu})^p.\label{1}
\end{equation}%
$F_{\mu\nu}=\partial_\mu A_\nu-\partial_\nu A_\mu$ is the electromagnetic tensor, where $A_\mu$ represents the electromagnetic potential. $p$ is a positive integer and the above action recovers the standard Maxwell action when $p=1$. It can be proved that the energy-momentum tensor can be traceless provided $n=4p$. Then the action of $R+f(R)$ gravity coupled to conformally invariant Maxwell field in $n$-dimensional spacetime reads
\begin{equation}
S=\int_\mathcal{M} d^nx\sqrt{-g}[R+f(R)-(F_{\mu\nu}F^{\mu\nu})^p],\label{2}
\end{equation}%
where $f(R)$ is an arbitrary function of scalar curvature $R$.

The corresponding $n$-dimensional black hole metric has been derived in Ref.~\cite{Sheykhi} as follow
\begin{equation}
ds^2=-N(r)dt^2+\frac{dr^2}{N(r)}+r^2 d\Omega^2_{n-2},\label{3}
\end{equation}%
where
\begin{equation}
N(r)=1-\frac{2m}{r^{n-3}}+\frac{q^2}{r^{n-2}}\times\frac{(-2q^2)^{(n-4)/4}}{1+f'(R_0)}-\frac{R_0}{n(n-1)}r^2.\label{4}
\end{equation}%
Note that the above black hole solutions hold for the dimensions which are multiples of four due to the restriction $n=4p$ ensuring the traceless property of the energy-momentum tensor, i.e., $n=4,8,12,...$ . When $n=4$, the above solutions recover the solution in Ref.~\cite{Moon98}. As argued in Ref.~\cite{Sheykhi}, for the case $1+f'(R_0)>0$, there would be two inner and outer horizons, an extreme black hole or naked singularity due to different choices of parameters. However, for $1+f'(R_0)<0$, the conserved quantities such as mass would be negative, making this case nonphysical~\cite{Sheykhi}.

It has been pointed out in Ref.~\cite{Sheykhi} that the above solution is asymptotically AdS if one define $R_0=-n(n-1)/l^2$. But it differs from the higher-dimensional RN-AdS black hole for its electric charge term go as $r^{-(n-2)}$ while that of RN-AdS black holes go as $r^{-2(n-3)}$.

The mass $M$, entropy $S$, charge $Q$, electric potential $\Phi$, and Hawking temperature $T$ was obtained as follow~\cite{Sheykhi}
\begin{eqnarray}
M&=&\frac{(n-2)\Omega_{n-2}}{8\pi}m(1+f'(R_0)),\label{5}
\\
S&=&\frac{r_+^{n-2}\Omega_{n-2}}{4}(1+f'(R_0)),\label{6}
\\
Q&=&\frac{n(-2)^{(n-4)/4}q^{(n-2)/2}\Omega_{n-2}}{16\pi\sqrt{1+f'(R_0)}},\label{7}
\\
\Phi&=&\frac{q}{r_+}\sqrt{1+f'(R_0)},\label{8}
\\
T&=&\frac{(1+f'(R_0))\times[2r_+^2(n-1)+2l^2(n-3)]+(-2q^2)^{n/4}r_+^{2-n}l^2}{8\pi l^2 r_+(1+f'(R_0))},\label{9}
\end{eqnarray}%
where $\Omega_{n-2}$ denotes the volume of the unit (n-2)-sphere. Note that we have corrected the typo error in the expression of $T$ by adding the coefficient $8$ in its denominator so that it can recover the Hawking temperature of four-dimensional black holes~\cite{Moon98} when $n=4$.

It was proved that the above physical quantities satisfy the first law of black hole thermodynamics~\cite{Sheykhi}
\begin{equation}
dM=TdS+\Phi dQ.\label{10}
\end{equation}%

\section{Extended phase space thermodynamics of higher-dimensional f(R) black holes}
\label{Sec3}
In this section, we introduce the extended phase space where the cosmological constant is identified as the thermodynamic pressure while the conjugate quantity is regarded as the thermodynamic volume.

Here we adopt the following definition of pressure that is commonly used in former literatures of extended phase space
\begin{equation}
P=\frac{-\Lambda}{8\pi}=\frac{(n-1)(n-2)}{16\pi l^2}.\label{11}
\end{equation}%
Note that we have utilized the expression of the cosmological constant in the $n$-dimensional spacetime $\Lambda=-\frac{(n-1)(n-2)}{2l^2}$. Comparing it with the definition of $R_0$, one can find the relation between $\Lambda$ and $R_0$ as
\begin{equation}
R_0=\frac{2n\Lambda}{n-2},\label{12}
\end{equation}%
which recovers the relation $R_0=4\Lambda$ when $n=4$.

Utilizing Eqs. (\ref{4}), (\ref{5}) and (\ref{11}), one can derive
 \begin{equation}
V=\left(\frac{\partial M}{\partial P}\right)_{S,Q}=\frac{(1+f'(R_0))r_+^{n-1}\Omega_{n-2}}{n-1}.\label{13}
\end{equation}%
Comparing it with the thermodynamic volume of $n$-dimensional RN-AdS black holes~\cite{Gunasekaran}, one can find that it gains an extra factor $1+f'(R_0)$ due to the effect of $f(R)$ gravity.

Utilizing Eqs. (\ref{5}), (\ref{6}), (\ref{7}), (\ref{8}), (\ref{9}), (\ref{11}) and (\ref{13}), the first law of black hole thermodynamics and the Smarr relation in the extended phase space can be obtained as
\begin{eqnarray}
dM&=&TdS+\Phi dQ+VdP,\label{14}
\\
M&=&\frac{n-2}{n-3}TS+\frac{(n-2)^2}{n(n-3)}\Phi Q-\frac{2}{n-3}VP.\label{15}
\end{eqnarray}%

When $n=4$, the above result reduces to $M=2TS+\Phi Q-2VP$, coinciding with the Smarr relation of RN-AdS black holes.

Utilizing Eqs. (\ref{9}) and (\ref{11}), the equation of state can be derived as
 \begin{equation}
P=\frac{(n-2)T}{4r_+}-\frac{(n-3)(n-2)}{16\pi r_+^2}-\frac{(n-2)2^{n/4}(-q^2)^{n/4}}{32(1+f'(R_0))\pi r_+^n}.\label{16}
\end{equation}%

Identifying the specific volume $v$ as $v=\frac{4r_+}{(n-2)}$, the above equation can be rewritten as
 \begin{equation}
P=\frac{T}{v}-\frac{(n-3)}{(n-2)\pi v^2}-\frac{2^{9n/4}(-q^2)^{n/4}}{32(1+f'(R_0))\pi (n-2)^{n-1}v^n}.\label{17}
\end{equation}%

The possible critical point can be defined by
\begin{eqnarray}
\left(\frac{\partial P}{\partial v}\right)_{T=T_c, v=v_c}&=&0,\label{18}
\\
\left(\frac{\partial^2 P}{\partial v^2}\right)_{T=T_c, v=v_c}&=&0.\label{19}
\end{eqnarray}%

Substituting Eq. (\ref{17}) into the above two equations, one can easily get
\begin{eqnarray}
-\frac{T_c}{v_c^2}+\frac{2(n-3)}{(n-2)\pi v_c^3}+\frac{2^{9n/4}(-q^2)^{n/4}n}{32(1+f'(R_0))\pi (n-2)^{n-1}v_c^{n+1}}&=&0,\label{20}
\\
\frac{2T_c}{v_c^3}-\frac{6(n-3)}{(n-2)\pi v_c^4}-\frac{2^{9n/4}(-q^2)^{n/4}n(n+1)}{32(1+f'(R_0))\pi (n-2)^{n-1}v_c^{n+2}}&=&0.\label{21}
\end{eqnarray}%
Simplifying the equation that (\ref{20})$\times (n+1)+$(\ref{21})$\times v=0$, one can derive the relation between the possible critical specific volume and critical temperature as
\begin{equation}
T_c=\frac{2(n-3)}{(n-1)\pi v_c}.\label{22}
\end{equation}%
Substituting Eq. (\ref{22}) into Eq. (\ref{20}), one can obtain
\begin{equation}
v_c^{n-2}=\frac{-n(n-1)2^{\frac{9n-24}{4}}(-q^2)^{n/4}}{(1+f'(R_0))(n-3)(n-2)^{n-2}}.\label{23}
\end{equation}%
Substituting Eqs. (\ref{22}) and (\ref{23}) into Eq. (\ref{17}), the possible critical pressure can be derived as
\begin{equation}
P=\frac{n-3}{n\pi}\left[\frac{-n(n-1)2^{\frac{9n-24}{4}}(-q^2)^{n/4}}{(1+f'(R_0))(n-3)(n-2)^{n-2}}\right]^{\frac{-2}{n-2}}.\label{24}
\end{equation}%
Keeping in mind the restrictions that $n=4p$ and $1+f'(R_0)>0$, we can draw the conclusion from Eq. (\ref{23}) that meaningful critical specific volume only exists when $p$ is odd. Because $v_c^{n-2}<0$ when $p$ is even.

The ratio $\frac{P_cv_c}{T_c}$ can be calculated as
\begin{equation}
\frac{P_cv_c}{T_c}=\frac{n-1}{2n}.\label{25}
\end{equation}%
The above result differs from that of higher dimensional charged AdS black holes in Einstein gravity. However, when $n=4$, the ratio $\frac{P_cv_c}{T_c}$ reduces to $3/8$, which is the same as that of four-dimensional RN-AdS black holes, implying that $f(R)$ gravity does not influence the ratio $\frac{P_cv_c}{T_c}$. So the result of Eq. (\ref{25}) may be attributed to the effect of conformally invariant Maxwell field. Note that Ref.~\cite{Chen} argued
that for four dimensional $f(R)$ black holes this ratio is influenced by the $f(R)$ correction. This inconsistency with our results can be attributed to the fact that the pressure in Ref.~\cite{Chen} was defined as $\frac{-bR_0}{32\pi}$ and different from the traditional definition by adding the factor $b=1+1+f'(R_0)$.

In the extended phase space, the mass of black holes should be interpreted as enthalpy. Then the Gibbs free energy can be derived as
\begin{equation}
G=H-TS=M-TS.\label{26}
\end{equation}%
Utilizing Eqs. (\ref{5}), (\ref{6}), (\ref{9}) and (\ref{11}), one can obtain
\begin{equation}
G=\Omega_{n-2}[\frac{(1+f'(R_0))r_+^{n-3}}{16\pi}-\frac{(1+f'(R_0))Pr_+^{n-1}}{(n-1)(n-2)}-\frac{(n-1)2^{\frac{n-20}{4}}(-q^2)^{n/4}}{\pi r_+}].\label{27}
\end{equation}%

Some specific cases will be discussed as follows.

\textbf{Case 1: $p=1$}

When $p=1$, one can obtain
\begin{equation}
T_c=\frac{\sqrt{6(1+f'(R_0))}}{18\pi q},\;v_c=\frac{2\sqrt{6}\,q}{\sqrt{1+f'(R_0)}},\;P_c=\frac{1+f'(R_0)}{96\pi q^2},\label{28}
\end{equation}%
recovering the critical Hawking temperature and critical volume of four-dimensional $f(R)$ AdS black holes~\cite{Chen}. However, the numerator of our critical pressure is $1+f'(R_0)$ instead of $(1+f'(R_0))^2$ since due to different definitions of $P$.

The Gibbs free energy reduces to
\begin{equation}
G=\frac{(1+f'(R_0))r_+}{4}-\frac{2(1+f'(R_0)) P \pi r_+^3}{3}+\frac{3q^2}{4r_+}, \label{29}
\end{equation}%
which matches Eq.(24) in Ref.~\cite{Chen} except the extra factor $(1+f'(R_0))$ in the second term due to different definitions of $P$.

\textbf{Case 2: $p=2$}

When $p=2$, Eq. (\ref{23}) reduces to
\begin{equation}
v_c^{6}=\frac{-3584q^4}{3645(1+f'(R_0))}<0,\label{30}
\end{equation}%
from which one can not obtain positive real root for the critical specific volume. So there is no meaningful critical point for eight-dimensional $f(R)$ AdS black holes. This result is quite different from four-dimensional f(R) AdS black holes. It may be attributed to the combined effect of $f(R)$ gravity and conformally invariant Maxwell field.

\textbf{Case 3: $p=3$}

When $p=3$, the critical quantities can be calculated as
\begin{eqnarray}
v_c^{10}&=&\frac{11\times2^{13}q^6}{3\times5^{10}(1+f'(R_0))},\label{31}
\\
T_c&=&\frac{18}{11\pi v_c},\label{32}
\\
P_c&=&\frac{3}{4\pi v_c^2},\label{33}
\\
\frac{P_cv_c}{T_c}&=&\frac{11}{24}.\label{34}
\end{eqnarray}%
From above, one can safely draw the conclusion that the critical physical quantities depend on both the parameter $q$ and the term $f'(R_0)$ which reflects
the effect of $f(R)$ gravity. With the increasing of $q$, both $T_c$ and $P_c$ decrease while $v_c$ increases. With the increasing of $f'(R_0)$, both $T_c$ and $P_c$ increase while $v_c$ decreases. However, the ratio $\frac{P_cv_c}{T_c}$ keeps constant. Table \ref{tb1} lists the relevant critical quantities for different choices of parameters.

\begin{table}[!h]
\tabcolsep 0pt
\caption{Critical quantities for $p=3$}
\vspace*{-12pt}
\begin{center}
\def\temptablewidth{0.5\textwidth}
{\rule{\temptablewidth}{1pt}}
\begin{tabular*}{\temptablewidth}{@{\extracolsep{\fill}}ccccc}
$q$ & $f'(R_0)$ & $v_c$ &$T_c$ &$P_c$ \\   \hline
    1    & 0.5     & $0.538502$        &        0.967259 & 0.823261  \\
     1  & 0.2     & $0.550653$        &        0.945915 & 0.787327   \\
 1     &  0     & $0.560785$        &        0.928825 & 0.759135   \\
    1    & -0.2     & $0.573439$        &        0.908328 & 0.726001   \\
       1     & -0.5     & $0.601034$        &        0.866624 & 0.660866   \\
            2     & 0.2     & $0.834634$        &        0.624071 & 0.342704  \\
                3    & 0.2     & $1.06451$        &        0.489304 & 0.210673
       \end{tabular*}
       {\rule{\temptablewidth}{1pt}}
       \end{center}
       \label{tb1}
       \end{table}

Utilizing Eqs. (\ref{17}) and (\ref{27}), one can derive both the equation of state and the Gibbs free energy for the case $p=3$ as
\begin{eqnarray}
P&=&\frac{T}{v}-\frac{9}{10\pi v^2}+\frac{2^{27}q^6}{32\times 10^{11}(1+f'(R_0))\pi v^{12}},\label{35}\\
G&=&0.412314(1+f'(R_0))r_+^{9}-0.18841(1+f'(R_0))Pr_+^{11}+\frac{18.1418q^6}{r_+}.\label{36}
\end{eqnarray}%

\begin{figure*}
\centerline{\subfigure[]{\label{1a}
\includegraphics[width=8cm,height=6cm]{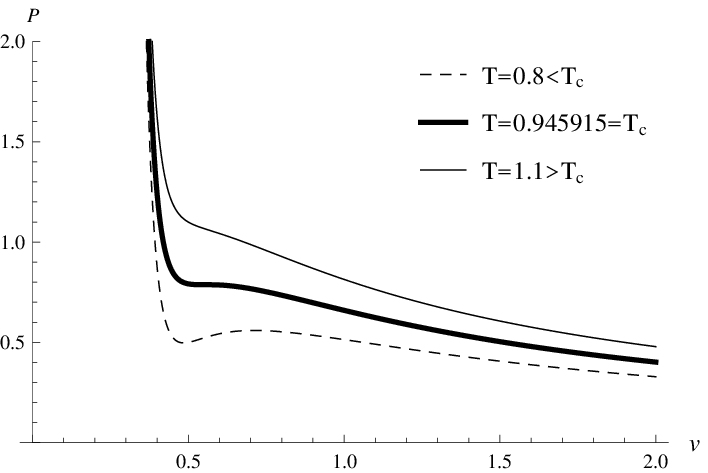}}
\subfigure[]{\label{1b}
\includegraphics[width=8cm,height=6cm]{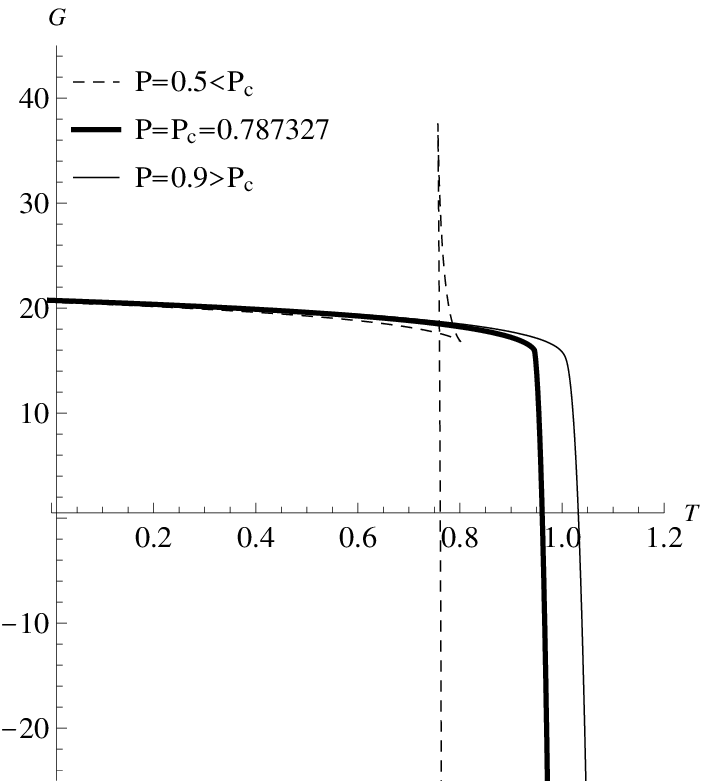}}}
 \caption{(a) $P$ vs. $v$ for $p=3,q=1,f'(R_0)=0.2$ (b) $G$ vs. $T$ for $p=3,q=1,f'(R_0)=0.2$} \label{fg1}
\end{figure*}

To gain an intuitive understanding, we plot the $P-v$ graph in Fig. \ref{1a} and Gibbs free energy graph in Fig. \ref{1b} for the case $q=1,f'(R_0)=0.2$. The isotherm can be divided into three branches when the temperature is lower than the critical temperature. Both the large radius branch and the small radius branch are stable while the medium radius branch is unstable. Phase transition analogous to the van der Waals liquid-gas system take place between the small black hole and the large black hole. Classical swallow tail behavior, characteristic of first order phase transition, can also be observed in the Gibbs free energy graph.

\section{Critical exponents of higher-dimensional f(R) black holes}
\label{Sec4}
To investigate the critical behavior near the critical point, we would like to calculate the relevant critical exponents as follows
\begin{eqnarray}
C_v&\propto&|t|^{-\alpha},\label{37}\\
\eta&\propto&|t|^{\beta},\label{38}\\
\kappa_T&\propto&|t|^{-\gamma},\label{39}\\
|P-P_c|&\propto&|v-v_c|^{\delta}.\label{40}
\end{eqnarray}
From the definitions above, one can see clearly that $\alpha$ and $\gamma$ characterize the behavior of fixed volume specific heat $C_v$ and isothermal compressibility coefficient $\kappa_T$ respectively while $\beta$ reflects the behavior of the order parameter $\eta$. The last critical exponent $\delta$ describes the behavior of the critical isotherm.

The fixed volume specific heat $C_v$ equals to zero, because the entropy $S$ is independent of the Hawking temperature $T$, which can be witnessed from Eq. (\ref{6}). Then the conclusion that $\alpha=0$ can be drawn.

To characterize to what extent the physical quantities approach the critical point, it would be convenient to introduce the notations below
\begin{equation}
t=\frac{T}{T_c}-1,\;\;\epsilon=\frac{v}{v_c}-1,\;\;p=\frac{P}{P_c}.\label{41}
\end{equation}
Utilizing Eq. (\ref{41}), Eq. (\ref{17}) can be expanded as follow near the critical point
  \begin{equation}
p=1+a_{10}t+a_{01}\epsilon+a_{11}t\epsilon+a_{02}\epsilon^2+a_{03}\epsilon^3+O(t\epsilon^2,\epsilon^4).\label{42}
\end{equation}
where
\begin{eqnarray}
a_{01}&=&a_{02}=0,\label{43}\\
a_{10}&=&\frac{T_c}{P_cv_c},\label{44}\\
a_{11}&=&-\frac{T_c}{P_cv_c},\label{45}\\
a_{03}&=&-\frac{T_c}{P_cv_c}+\frac{4(n-3)}{(n-2)\pi v_c^2P_c}+\frac{2^{9n/4}n(n+1)(n+2)(-q^2)^{n/4}}{192(n-2)^{n-1}(1+f'(R_0))\pi v_c^{n}P_c}.\label{46}
\end{eqnarray}

Denoting the large black hole and small black hole with the subscripts $l$ and $s$ respectively, one can obtain
 \begin{equation}
1+a_{10}t+a_{11}t\epsilon_l+a_{03}\epsilon_l^3=1+a_{10}t+a_{11}t\epsilon_s+a_{03}\epsilon_s^3.\label{47}
\end{equation}
The above equation is based on the fact that during the phase transition the pressure keeps constant.

According to Maxwell's equal area law,
 \begin{equation}
\int^{\epsilon_s}_{\epsilon_l}\epsilon \frac{dp}{d\epsilon}d \epsilon=0.\label{48}
\end{equation}
Utilizing Eq. (\ref{42}), one can easily obtain
\begin{equation}
\frac{dp}{d\epsilon}=a_{11}t+3a_{03}\epsilon^2,\label{49}
\end{equation}
Substituting it into Eq. (\ref{48}), one can get
 \begin{equation}
 a_{11}t(\epsilon^2_s-\epsilon^2_l)+\frac{3}{2} a_{03}(\epsilon^4_s-\epsilon^4_l)=0.\label{50}
\end{equation}
From Eqs. (\ref{32}) and (\ref{35}), one can obtain
\begin{equation}
\epsilon_l=-\epsilon_s=\sqrt{\frac{-a_{11}t}{a_{03}}}.\label{51}
\end{equation}
So
\begin{equation}
\eta=v_l-v_s=v_c(\epsilon_l-\epsilon_s)=2v_c\epsilon_l\propto\sqrt{-t},\label{52}
\end{equation}
Comparing it with the definition of $\beta$, one can conclude that $\beta=1/2$.

The isothermal compressibility coefficient $\kappa_T$ can be calculated as
\begin{equation}
\kappa_T=\left.-\frac{1}{v}\frac{\partial v}{\partial P}\right|_{v_c}\propto \left.-\frac{1}{\frac{\partial p}{\partial \epsilon}}\right|_{\epsilon=0}=-\frac{1}{a_{11}t},\label{53}
\end{equation}
yielding $\gamma=1$.

 Substituting $t=0$ into Eq. (\ref{42}), one can get
\begin{equation}
p-1=a_{03}\epsilon^3,\label{54}
\end{equation}
Comparing it with the definition of $\delta$, one can draw the conclusion that $\delta=3$.

Our results of critical exponents are exactly the same as those obtained before for other AdS black holes in former literatures~\cite{Kubiznak,Gunasekaran,Chen,Hendi}, implying that neither $f(R)$ gravity nor conformally invariant Maxwell field influence the critical exponents. This can be attributed to the mean field theory.

\section{Discussions on the constraint on the function $f(R)$}
\label{Sec5}
Till now, we have accomplished the research goal of generalizing the thermodynamics of higher-dimensional $f(R)$ black holes to the extended phase space. It is shown that the critical physical quantities depend on the term $f'(R_0)$, showing the effect of $f(R)$ gravity. Note that the investigated black hole solution depends on the form of the function $f(R)$. Here, we will show that how our results put constraint on the form of the function $f(R)$.

From Eq. (\ref{4}), one can see clearly that the term $1+f'(R_0)$ appears in the function $N(r)$, thus influencing the black hole metric. The black hole solution is asymptotically AdS if one define $R_0=-n(n-1)/l^2$. With this definition, $R_0$ is proportional to cosmological constant, which can be witnessed from Eq. (\ref{12}). So the analyses in this paper contain the underlying assumption that $R_0$ is negative. It can be viewed as the first constraint that our results put on the function $f(R)$ since $R_0$ can be non-negative in general.

The second constraint comes from the physical consideration. As reviewed in Sec. \ref{Sec2}, there would be two inner and outer horizons, an extreme black hole or naked singularity due to different choices of parameters for the case $1+f'(R_0)>0$. However, for $1+f'(R_0)<0$, the conserved quantities such as mass would be negative, making this case nonphysical~\cite{Sheykhi}.

An example will be presented below for one to see how these two conditions constraint the specific form of $f(R)$. Considering a simple case that $f(R)=a_0R^2+b_0R^3$, one can easily derive
\begin{equation}
1+f'(R_0)=1+2a_0R_0+3b_0R_0^2.\label{55}
\end{equation}
To obtain $R_0$, we can seek help from the important relation presented in the former literature. Considering the trace of the equation of motion, Refs. ~\cite{Dombriz, Sheykhi} derived that
\begin{equation}
R_0(1+f'(R_0))-\frac{n}{2}(R_0+f(R_0))=0.\label{56}
\end{equation}
Substituting $f(R)=a_0R^2+b_0R^3$ into the above relation, one can derive
\begin{equation}
-\frac{R_0}{2}[b_0(n-6)R_0^2+a_0(n-4)R_0+n-2]=0.\label{57}
\end{equation}
Solving Eq. (\ref{57}), one can obtain its nonzero roots as
\begin{equation}
R_0=\frac{(4-n)a_0\mp \sqrt{(n-4)^2a_0^2-4(n-2)(n-6)b_0}}{2b_0(n-6)}.\label{58}
\end{equation}
Substituting Eq. (\ref{58}) into Eq. (\ref{55}), one can further get
\begin{equation}
1+f'(R_0)=\frac{n\left[(n-4)a_0^2-4b_0(n-6)\mp a_0\sqrt{(n-4)^2a_0^2-4(n-2)(n-6)b_0}\right]}{2b_0(n-6)^2}.\label{59}
\end{equation}
Solving both the inequalities $R_0<0$ and $1+f'(R_0)>0$, one can figure out the restriction on the model parameter $a_0$ and $b_0$.

Due to the restriction that $n=4p$, $n$ can be chosen as $4, 8, 12$.
When $n=4$, Eqs. (\ref{58}) and (\ref{59}) reduce to
\begin{eqnarray}
R_0&=&-\frac{1}{\sqrt{b_0}},\label{60}
\\
1+f'(R_0)&=&\frac{4a_0\sqrt{b_0}+8b_0}{2b_0}.\label{61}
\end{eqnarray}
Note that we have omitted the positive root of $R_0$. And one can easily obtain the restriction on $a_0$ and $b_0$ to be $b_0>0, a_0>-2\sqrt{b_0}$. Similarly, one can find the constraint for the cases $n=8,12$ and other models of $f(R)$. 

\section{Conclusions}
\label{Sec6}
In this paper, we generalize the former research on the thermodynamics of higher-dimensional $f(R)$ black holes to the extended phase space, where the cosmological constant is viewed as a variable. We introduce the traditional definition of thermodynamic pressure as $P=-\frac{\Lambda}{8\pi}$ and derive its conjugate quantity as thermodynamic volume. It is shown that the thermodynamic volume gains an extra factor $1+f'(R_0)$ due to the effect of $f(R)$ gravity. We obtain both the first law of black hole thermodynamics and Smarr relation in the extended phase space.

With the definition of thermodynamic pressure, we also derive the equation of state. By solving the critical condition equations, we obtain the analytic expressions for the possible critical Hawking temperature, pressure and specific volume. We also list the numerical results of critical quantities for different choices of parameters. It is proved that meaningful critical specific volume only exists when $p$ is odd. The critical physical quantities depend on both the parameter $q$ and the term $f'(R_0)$. With the increasing of $q$, both $T_c$ and $P_c$ decrease while $v_c$ increases. With the increasing of $f'(R_0)$, both $T_c$ and $P_c$ increase while $v_c$ decreases. However, the ratio $\frac{P_cv_c}{T_c}$ keeps constant. It is shown that the ratio $P_cv_c/T_c$ differs from that of higher dimensional charged AdS black holes in Einstein gravity. However, when $n=4$, the ratio $\frac{P_cv_c}{T_c}$ reduces to $3/8$, which is the same as that of four-dimensional RN-AdS black holes, implying that $f(R)$ gravity does not influence the ratio $\frac{P_cv_c}{T_c}$. So our results may be attributed to the effect of conformally invariant Maxwell field. Note that Ref.~\cite{Chen} argued that for four dimensional $f(R)$ black holes this ratio is influenced by the $f(R)$ correction. This inconsistency with our results can be attributed to the fact that the pressure in Ref.~\cite{Chen} was defined as $\frac{-bR_0}{32\pi}$ and different from the traditional definition by adding the factor $b=1+1+f'(R_0)$.

To probe the phase transition, we derive the explicit expression of the Gibbs free energy. $P-v$ graph and Gibbs free energy graph are also plotted. The isotherm can be divided into three branches when the temperature is lower than the critical temperature. Both the large radius branch and the small radius branch are stable while the medium radius branch is unstable. Phase transition analogous to the van der Waals liquid-gas system take place between the small black hole and the large black hole. Classical swallow tail behavior, characteristic of first order phase transition, can also be observed in the Gibbs free energy graph. To investigate the critical behavior near the critical point, critical exponents are calculated. It is shown that these exponents are exactly the same as those of other AdS black holes reported in literature before, implying that neither $f(R)$ gravity nor conformally invariant Maxwell field influence the critical exponents. This can be attributed to the mean field theory. Since the investigated black hole solution depends on the form of the function $f(R)$, we discuss in detail how our results put constraint on the form of the function $f(R)$ and we also present a simple example. 

Our research further deepen the understanding of relation between AdS black holes and liquid-gas systems. On the other hand, our research also discover some unique characteristics that may be attributed to the combined effect of $f(R)$ gravity and conformally invariant Maxwell field. Ref.~\cite{Hendi3} discovered an interesting relation between the solutions of a class of pure $f(R)$ gravity and those of Einstein conformally invariant Maxwell source. This may be a promising direction to probe the rich physics behind the interesting critical phenomena that we reported in this paper. Further investigation is called for.

 \section*{Acknowledgements}
We would like to express our sincere gratitude to the anonymous referee whose deep insight has improved the quality of this paper greatly. This research is supported by Guangdong Natural Science Foundation (Grant Nos.2016A030307051, 2016A030310363, 2015A030313789) and Department of Education of Guangdong Province of China(Grant No.2014KQNCX191). It is also supported by \textquotedblleft Thousand Hundred Ten\textquotedblright \,Project of Guangdong Province.

\end{document}